\documentclass[preprint,aps,showpacs,preprintnumbers,amsmath,amssymb]{revtex4}

\usepackage{graphicx}
\usepackage{dcolumn}
\usepackage{bm}

\newcommand{\beq}{\begin{equation}}
\newcommand{\eq}{\end{equation}}
\newcommand{\bea}{\begin{eqnarray}\displaystyle}
\newcommand{\ea}{\end{eqnarray}}
\newcommand{\etap}{\eta_P}
\newcommand{\etat}{\tilde{\eta}}

\newcommand{\Bl}{\Bigl}
\newcommand{\Br}{\Bigr}
\newcommand{\bl}{\bigl}
\newcommand{\br}{\bigr}

\begin{document}

\preprint{}

\title{The Effects of Quantum Deformations \\ on the Spectrum of Cosmological
Perturbations}
\author{Sera Cremonini}

 \email{sera@het.brown.edu}
\affiliation{Department of Physics, Brown University \\ Providence,
RI 02912, USA}

\date{\today}

\begin{abstract}
We consider a quantum deformation of the wave equation on a cosmological
background as a toy-model for possible trans-Planckian effects.
We compute the power spectrum of scalar and tensor fluctuations
for power-law inflation, and find a noticeable deviation from the standard result.
We consider de Sitter inflation as a special case, and find that the resulting power spectrum is scale invariant.
For both inflationary scenarios the departure from the standard
spectrum is sensitive to the size of the deformation parameter.
A modulation in the power spectrum appears to be a generic feature of the model.
\end{abstract}

\pacs{98.80Cq}
\maketitle

\section{Introduction}

Recently a lot of attention has been devoted to the possibility
that Planck scale physics might leave an imprint on the cosmic
microwave background (CMB).
The temperature anisotropies of the CMB in fact arise from
quantum fluctuations generated in the early universe
and stretched by inflation to observable scales.
Thus, length scales which are now of cosmological size could have been
below the Planck length at the beginning of inflation,
and could therefore carry the imprint of trans-Planckian physics.
Within the standard theory of cosmological perturbations, fluctuations are
assumed to originate in the infinite past with an infinitely small size.
However, near the Planck scale the structure of space-time is
believed to be modified by quantum gravitational effects,
with the Planck length playing the role of a physical cutoff. Such modifications could
affect the evolution of cosmological perturbations, possibly even
changing the predictions of standard inflation.

Several groups have considered models for physics above the Planck scale,
and computed the resulting corrections to the standard spectrum of fluctuations.
These approaches have included nonlinear dispersion relations
 \cite{Martin:2000xs,Brandenberger:2000wr,Amelino-Camelia:1999zc,Niemeyer,Mersini:2001su,Starobinsky:2001kn,BM1,Brandenberger:2002hs,NP2,LLMU,BJM},
 short-distance modifications to quantum-mechanical commutation
relations \cite{Kempf,CGS,EGKS,KempfN,HS}, models based on noncommutative
geometry \cite{Lizzi:2002ib,Brandenberger:2002nq,Huang:2003zp},
non-standard initial conditions
 \cite{Danielsson:2002kx,Danielsson:2002qh,Danielsson:2002mb,Goldstein:2002fc,Alberghi:2003am,Armendariz-Picon:2003gd}
 and effective field theory techniques \cite{Shenker}.
Also, more recent work has been based on `minimal' models, where the specific nature
of trans-Planckian physics is not assumed, but rather described by the boundary conditions
imposed on the modes at the cutoff scale \cite{Danielsson:2002kx,Easther:2002xe,Niemeyer:2002kh,Martin:2003kp}.
For a detailed review, see \cite{Martin:2003kp}.

What emerges from these proposals is the possibility of obtaining
substantial corrections to the standard CMB power spectrum,
showing that inflation can indeed carry the signature of
physics close to the string (Planck) scale.
However, it is also apparent that the conclusions reached are extremely
sensitive to the specific short-distance modifications employed and to the choices of initial
conditions.
Due to the lack of a fundamental theory of quantum gravity, many
of the models that have been brought forth are based on {\em ad hoc} modifications to
short-distance physics. Thus, it is not surprising that they
yield different predictions.
Still, until we have a more fundamental theory of gravity, a bottom-up approach to
modeling very high energy physics is valuable {\em per se}, and so is the
effort to construct viable cosmological scenarios that might be tested by looking at the CMB
temperature anisotropies.

Although slightly different, most of the models considered above
share the idea that the structure of space-time
is endowed with a fundamental length, usually taken to be the
Planck length $l_P=M_P^{-1}$, corresponding to a high-energy cutoff.
It is the presence of such a fundamental
scale that is responsible for curing the ultraviolet
behavior of the underlying theory, causing processes with energy higher than
the cutoff to be suppressed.
Also, many quantum gravity models share the notion of a `fuzzy' structure of space-time.
For instance, noncommutative geometry provides a natural framework for the formulation of field theories
that incorporate fundamental space-time uncertainties.
Quantum ($Q$-)deformations, also an example of noncommutativity, provide an additional
type of regularization scheme.
They are well motivated not only because they are a powerful candidate for regulating divergences,
but also because they preserve some of the symmetries of the underlying theory, in the
form of quantum symmetries \cite{Majid:1990gq}.
For an example of quantum groups in the context of Planck scale
physics and of an algebraic approach to quantum gravity see \cite{Majid:1996nt} and references
therein.

In this work we consider a nonlocal generalization of the wave
equation describing cosmological perturbations, in the form of a quantum
deformation. The deformation consists in replacing the standard wave operator
by its $Q$-deformed counterpart. The new operator is nonlocal by an amount
dictated by how much the deformation parameter deviates from its classical value.
For similar work in the context of a two-dimensional black hole
see \cite{Teschner:1999my}.

This paper is structured as follows.
We start in section I with a brief review of the standard theory of
cosmological perturbations. In the next section we find the solution to
the $Q$-deformed mode equation at scales higher than the cutoff $\Lambda$, and
match it to the standard solution valid below the cutoff scale.
We make a conservative choice for the initial conditions, taking a Bunch-Davies-like vacuum.
Section III gives the correction to the standard power spectrum for both tensor and scalar
fluctuations for the case of power-law inflation. Finally, in section IV we repeat the
calculations for the case of de Sitter inflation. We conclude by
discussing the implications of our results.

\section{Standard Inflation}
In this section we review very briefly the main features of the theory of cosmological
perturbations within the framework of standard inflation.
Power-law inflation in a spatially flat FRW universe is described by the metric
\beq
\label{metric}
ds^2=dt^2-a^2(t) d\vec{x}^2=a^2(\eta)\bl(d\eta^2-d\vec{x}^2\br),
\eq
where the scale factor is $a(t)=a_0 t^n$ or, in terms of
conformal time, $a(\eta)= \alpha_0 \eta^{\frac{n}{1-n}}$, with
$-\infty < \eta < 0$ and $\alpha_0\equiv \bl(a_0(1-n)\br)^{\frac{n}{1-n}}$.
Here $\alpha_0$ has dimensions of length.
The $n \rightarrow \infty$ limit of power-law inflation yields the de
Sitter background, where
$a(t)=e^{Ht}$ or $a(\eta)=-\frac{1}{H\eta}$, and
$H \equiv \frac{\dot{a}}{a}$ is a constant.

The two types of metric perturbations which are of interest in the
early universe are scalar and tensor fluctuations (see \cite{Mukhanov:1990me} for a review).
We start by considering the case of power-law inflation.
In Fourier space, the tensor perturbations $u_k$ obey the mode equation
\beq
\label{tenseq}
u^{\prime\prime}_k + \bl( k^2 - \frac{a^{\prime\prime}}{a} \br) u_k = 0,
\eq
where $k$ is the mode's comoving momentum and primes indicate derivatives with respect to conformal time.
Notice that this is the equation describing a parametric oscillator with a
time-dependent frequency.

Scalar perturbations couple to matter, and are responsible for the
large scale structure of the universe.
The mode equation governing the evolution of scalar fluctuations $v_k$ has a similar
structure,
\beq
\label{scalareq}
v^{\prime\prime}_k + \bl( k^2 - \frac{z^{\prime\prime}}{z} \br) v_k = 0,
\eq
with $a$ replaced by $z=a\frac{\phi_0^{\prime}}{{\cal{H}}}$.
Here ${\cal{H}} \equiv \frac{a^{\prime}}{a}$ and
$\phi_0$ denotes the background value of the scalar matter field.
In the case of power-law inflation $\phi_0^{\prime}$ and
${\cal{H}}$ are both proportional to the same power of $\eta$,
so that $\frac{z^{\prime\prime}}{z}=\frac{a^{\prime\prime}}{a}$.
Thus, scalar and tensor fluctuations evolve in the same way, and
the difference between them becomes manifest only in the final normalization
of the power spectrum.
For gravitational modes, the power spectrum is given by
\beq
\label{gravpowersp}
P_g(k) = \frac{k^3}{2\pi^2}\frac{|u_k|^2}{a^2}
\eq
while for scalar perturbations it is given by
\beq
\label{scalarpowersp}
P_s(k) = \frac{k^3}{2\pi^2}\frac{|v_k|^2}{z^2}.
\eq
Using the Friedman equations it can be shown that
$z = a\frac{\phi_0^{\prime}}{{\cal{H}}} \sim
\frac{a}{\sqrt{n}}$ for
power-law inflation.
This gives
\beq
\label{bothpowersp}
P_s(k) = \Bl(\frac{{\cal{H}}}{\phi_0^{\prime}}\Br)^2 P_g(k) \sim n P_g(k).
\eq
The case of de Sitter inflation is more subtle; while the
derivation of (\ref{gravpowersp}) still holds, that of
(\ref{scalarpowersp}) breaks down, and the spectrum of scalar
fluctuations is singular. Thus, we will only consider the spectrum
of tensor fluctuations for the de Sitter background. For a recent discussion of this point,
see \cite{Martin:2003kp}.

\section{Q-deformations and cosmological perturbations}

In this work we model the effects of quantum gravity on scales
larger than some cutoff energy $\Lambda$ by performing a
simple $Q$-deformation of the standard mode equation governing scalar and tensor
perturbations. We replace the usual derivatives appearing
in the mode equation by a finite difference operator,
\beq
\partial_r f(r)\longrightarrow D_r f(r),\ \ D_r f(r)\equiv
\frac{f(Qr)-f(r)}{Qr-r},
\eq
where $Q \: \epsilon \: \mathbb{R}$ and $Q\neq0,1$ is the parameter that controls the deformation.
The finite difference operator reduces to the standard derivative
when the deformation parameter is set equal to unity.
Some of the properties of $D_r$ which will be used throughout this paper are
\bea
D_r \Bl(f_1(r)f_2(r)\Br) &=& \Bl(D_r f_1(r)\Br)f_2(Qr) + f_1(r)D_r f_2(r), \\
D_r r^n &=& [n] r^{n-1}, \ \ \text {where} \ \ [n]\equiv \frac{Q^n-1}{Q-1}, \\
D_r e_Q^{ar} &=& ae_Q^{ar} \ \ \text{with} \ \
e_Q^{ar}=\sum_{n=0}^\infty \frac{a^n}{[n]!}r^n, \ \ [n]!\equiv [n][n-1]\dots[1].
\ea
Although one can in principle deform all coordinates, we choose a
deformation which affects only conformal time, motivated by an
analogy with the study of quantum deformations in Anti de Sitter
(AdS) space. In fact, it has been suggested \cite{Jevicki:1998rr} that
$Q$-deformations should act only on the radial coordinate of AdS space,
which corresponds to the time coordinate in de Sitter space.
Thus, the deformed equation of motion valid above the cutoff energy scale $\Lambda$ is
given by
\beq
\label{defeom1}
D^2_{\etat} u^{I}_{k} + \Bl(1 - \frac{1}{a}D^2_{\etat}a\Br) u^{I}_{k} =
0,\ \  \ \ \etat \equiv k\eta.
\eq
Notice that while we are explicitly considering tensor fluctuations, the
same analysis holds for scalar perturbations.
The difference between the two cases will become apparent in the final normalization
of the power spectrum.
Using
$\frac{1}{a}D^2_{\etat}a=\bl[\frac{n}{1-n}\br]\bl[\frac{n}{1-n}-1\br]\frac{1}{\etat^{2}}$,
eq. (\ref{defeom1}) simplifies to
\beq
\label{defeom2}
D^2_{\etat} u^{I}_{k} + \Bl(1 - \frac{C}{\etat^2}\Br) u^{I}_{k} = 0, \ \
 \ \ C\equiv \Bl[\frac{n}{1-n}\Br]\Bl[\frac{n}{1-n}-1\Br].
\eq
If we neglect the $\frac{C}{\etat^2}$ term,
$f_k(-k\eta) \sim e_Q^{-ik\eta}$ is a solution of the resulting $Q$-deformed equation of
motion.
Since in the trans-Planckian region $|k\eta|\gg 1$, we can construct
the solution to the full deformed mode equation by making the
expansion
\beq
\label{ansatz}
f_k(-k\eta) = e_Q^{-ik\eta} \sum_{j=0}^\infty b_j(-k\eta)^{-j}.
\eq
Plugging the ansatz (\ref{ansatz}) into (\ref{defeom2}) we find
\beq
f_k = e_Q^{-ik\eta}\Bl(b_0+\frac{b_2}{(k\eta)^2}+{\cal O}\bl(\frac{1}{(k\eta)^3}\br)  \Br) \ \
\text{with} \ \ b_2=\frac{Cb_0}{1-Q^{-4}}.
\eq

The general solution $u_k^I$ in region I, which we define to be above the cutoff scale $\Lambda$,
is a linear combination of positive and negative frequency modes,
\beq
\label{fulluI}
u_k^I = A_k^I  e_Q^{-ik\eta} \Bl(1 + \frac{1}{(k\eta)^2} \frac{C}{1-Q^{-4}} \Br) +
B_k^I  e_Q^{ik\eta} \Bl(1 + \frac{1}{(k\eta)^2} \frac{C}{1-Q^{-4}}  \Br),
\eq
where we set $b_0=1$.
$A_k^I$ and $B_k^I$ are determined by the choice of initial
conditions; by comparison with the Bunch-Davies (or adiabatic) vacuum, we choose
only the positive-frequency solution, and take $B_k^I=0$ and
$A_k^I=1$. However, more general choices of the coefficients are
also possible (\cite{Mottola:ar,Allen:ux}).

In region II, below the cutoff $\Lambda$, the exact solution to the mode equation
\beq
\label{modeeq}
\partial^2_{\eta} u^{II}_{k} + \Bl(k^2 - \frac{a^{\prime\prime}}{a}\Br) u^{II}_{k} = 0
\eq
is given by
\beq
\label{uII}
u^{II}_{k}=\sqrt{\eta} \Bl(A_k^{II} H_\nu(-k\eta)+B_k^{II} H^\ast_\nu(-k\eta)\Br),
\eq
where $\nu=\frac{3}{2} + \frac{1}{(n-1)}$ for power-law inflation and $\nu=\frac{3}{2}$ for de Sitter inflation, and
$H_\nu(-k\eta)$ is the Hankel function of the first kind.
The constants $A_k^{II},B_k^{II}$ satisfy $|A_k|^2-|B_k|^2=1$.
If we fix these coefficients by boundary conditions well inside the horizon,
we can follow the evolution of the mode (\ref{uII}) to the long wavelength limit,
where the power spectrum is defined, without having to match solutions at horizon crossing.
For $|k\eta| \gg 1$ the Hankel function $H_\nu(-k\eta)$ has the expansion
\beq \label{uIIasymp} H_\nu(-k\eta) = \sqrt{ \frac{2}{\pi
|k\eta|} } e^{-ik\eta -i(\frac{\pi\nu}{2}+\frac{\pi}{4})}\Bigl[1 -
\frac{i}{2k\eta}(\nu^2-\frac{1}{4}) -
\frac{(\nu^2-\frac{9}{4})(\nu^2-\frac{1}{4})}{8(k\eta)^2} + {\cal
O}\bl(\frac{1}{(k\eta)^3}\br)\Bigr].
\eq
Henceforth we will only keep terms up to order $\frac{1}{(k\eta)^2}$.

For $|k\eta| \gg 1$, the solution in region II is
\bea
\label{fulluII}
u^{II}_{k} & \sim & \sqrt{\frac{2}{\pi k}} \Biggl\{ A_k^{II} e^{-ik\eta
-i(\frac{\pi\nu}{2}+\frac{\pi}{4})} \Bigl[1 -
\frac{i(\nu^2-\frac{1}{4})}{2k\eta} -
\frac{(\nu^2-\frac{9}{4})(\nu^2-\frac{1}{4})}{8k^2\eta^2} \Bigr] + \nonumber \\ & & B_k^{II}
 e^{ik\eta +i(\frac{\pi\nu}{2}+\frac{\pi}{4})} \Bigl[1 +
\frac{i(\nu^2-\frac{1}{4})}{2k\eta} -
\frac{(\nu^2-\frac{9}{4})(\nu^2-\frac{1}{4})}{8k^2\eta^2} \Bigr] \Biggr\}  \nonumber \\
& \sim &  \sqrt{\frac{2}{\pi k}} \Biggl\{A_k^{II} e^{-ik\eta +i\phi} \Bl[1 -
\frac{iC_1}{k\eta} - \frac{C_2}{k^2\eta^2}\Br] +
B_k^{II} e^{ik\eta -i\phi} \Bl[1 + \frac{iC_1}{k\eta} - \frac{C_2}{k^2\eta^2}\Br]\Biggr\}
\ea
where $\phi \equiv -\frac{\pi\nu}{2}-\frac{\pi}{4}$, $C_1 \equiv \frac{\nu^2-\frac{1}{4}}{2}$
and $C_2 \equiv \frac{(\nu^2-\frac{9}{4})(\nu^2-\frac{1}{4})}{8}$.
Notice that while the standard solution $u^{II}_{k}$ contains a
term $\propto {k\eta}^{-1}$, $u^I_k$ does
not. Thus, the Q-deformed solution does not reduce to the standard
one in the limit $Q \rightarrow 1$.

The constants $A_k^{II}$, $B_k^{II}$ are found by matching
$u_k^I$ to the standard mode $u_k^{II}$ at the cutoff scale.
Taking $\Lambda=M_P$ and denoting by $\etap$ the conformal time at which
the mode crosses the cutoff scale, the matching conditions are
\bea
\label{matconds}
u_k^I\Big|_{\etap} &=& u_k^{II}\Big|_{\etap}, \\
D_\eta u_k^I\Big|_{\etap} &=& \partial_\eta u_k^{II}\Big|_{\etap}.
\ea
Solving the equations above we find
\bea
\label{coeffs}
A_k^{II} &=& e^{ik \etap - i\phi} e_Q^{-ik \etap} \sqrt{\frac{\pi k}{2}}
\Biggl\{1 + \frac{iC_1}{k\etap} + \frac{1}{(k\etap)^2} \Bl[\frac{C}{2(1-Q^{-2})} -C_1^2 +\frac{C_1}{2} +C_2\Br]\Biggr\},\\
B_k^{II} &=& e^{-ik \etap+i\phi} e_Q^{-ik \etap} \sqrt{\frac{\pi k}{2}}
\Biggl\{\frac{1}{(k\etap)^2}\Bl[\frac{C_1}{2}-\frac{C}{2(1+Q^{-2})}\Br]\Biggr\}.
\ea
Finally, we fix the normalization $N$ of the $u^{II}_k$ modes by requiring that
$|A_k|^2-|B_k|^2=1$,
\bea
|A_k^{II}|^2-|B_k^{II}|^2 &=& |N|^{-2} |e_Q^{-ik \etap}|^2 \frac{\pi k}{2}
 \Biggl \{ 1+ \frac{1}{(k\eta)^2} \Bl[ \frac{C}{(1-Q^{-2})} - C_1^2 + C_1 + 2C_2 \Br] \Biggr\} = 1 \nonumber \\
\Rightarrow |N|^2 &=& |e_Q^{-ik \etap}|^2 \frac{\pi k}{2}  \Biggl \{ 1+ \frac{1}{(k\eta)^2}
\Bl[ \frac{C}{(1-Q^{-2})} - C_1^2 + C_1 + 2C_2 \Br] \Biggr\}.
\ea

\section{Correction to the Standard Power Spectrum}
The power spectrum is calculated at late times, when $-k\eta\ll1$.
In this limit one has
\beq
H_{\nu}(-k\eta) \sim \frac{-i \: \Gamma(\nu) \:
  2^{\nu}}{\pi}(-k\eta)^{-\nu} \ \
\Rightarrow \ \  |u^{II}_k|=\sqrt{\eta} \ \frac{\Gamma(\nu) \: 2^{\nu}}{\pi}(-k\eta)^{-\nu}|A_k-B_k|.
\eq
Since, for both scalar and tensor fluctuations,
$P(k) \sim \frac{k^3}{a^2}|u^{II}_k|^2$,
one can express the power spectrum on super-horizon scales in the following simple way,
\beq
\label{powersp0}
P_{s,g}^{new}(k)=P_{s,g}^{st}(k) \ |A_k-B_k|^2,
\eq
where $A_k$ and $B_k$ are now understood to be properly normalized.
Thus, the effects of trans-Planckian physics are entirely contained in
the coefficients $A_k$ and $B_k$; the standard power spectrum
corresponds to having $B_k=0$ and $|A_k|^2=1$.

Using (\ref{powersp0}) we find that the lowest-order effect on the spectrum of
fluctuations due to the $Q$-deformation is given by
\beq
\label{powersp1}
P_{s,g}^{new}(k) = P_{s,g}^{st}(k) \: \Bl\{1-\frac{1}{(k\etap)^2}\bl[C_1-\frac{C}{(1+Q^{-2})}
\br]\cos(2k\etap-2\phi)\Br\},
\eq
where $P_{s,g}^{st}(k) \sim k^{\frac{2}{1-n}}$.
Recall that $\etap$ corresponds to the time when the physical momentum $p$ of the mode
equals the cutoff scale $M_P$,
\beq
p=\frac{k}{a(\etap)}=M_P \Rightarrow
\etap=\bl(\frac{k}{\alpha_0 M_P}\br)^{\frac{1-n}{n}}=\frac{1}{a_0(1-n)}{\Bl(\frac{k}{M_P}\Br)}^{\frac{1-n}{n}}.
\eq
Using this condition, the power spectrum can be rewritten as
\beq
\label{powersp2}
P_{s,g}^{new}(k) = P_{s,g}^{st}(k) \: \Biggl\{1-\Bl(\frac{1}{\alpha_0 M_P}\Br)^2
\Bl(\frac{k}{\alpha_0 M_P}\Br)^{-\frac{2}{n}}
\Bl(C_1-\frac{C}{1+Q^{-2}}\Br)
cos\Bl(\alpha_0 M_P \bl(\frac{k}{\alpha_0 M_P}\br)^{\frac{1}{n}}-2\phi \Br) \Biggr\}.
\eq
Notice that when $Q=1$ one obtains the standard result, $P_g^{new}(k) \rightarrow P_g^{st}(k)$.

In Fig. (\ref{fig1})
we plot the deviation from the standard power spectrum in the form of the ratio
\beq
\label{deviation}
R(k) = \frac{P_{s,g}^{new}(k)-P_{s,g}^{st}(k)}{P_{s,g}^{st}(k)}
\eq
for several values of $n$ and $Q$.
It is clear from the graphs that the oscillations around
$P_{s,g}^{st}(k)$ are substantial when $k$ is small, and become progressively
damped as $k$ grows larger, showing that the effects of the deformation are
more prominent for modes with small comoving momentum.
Moreover, the farther $Q$ is from the classical limit $Q=1$,
the larger the size of the correction.
Thus, a small enough $Q$ would increase the amplitude of the deviation from the standard
spectrum considerably.

We can compare the result (\ref{powersp2}) to the analysis of
\cite{Martin:2003kp}, where the modes are assumed to be created
at time $\eta_k$ when their wavelength equals the cutoff scale $M_C^{-1}$.
The authors of \cite{Martin:2003kp} introduced the quantity
\beq
\label{sigmak}
\sigma_k=\frac{\frac{n}{1-n}}{k\eta_k} \equiv
\frac{H(\eta_k)}{M_C}.
\eq
This can be written as
\beq
\label{sigmak2}
\sigma_k =\sigma_0 \bl(\frac{\eta_k}{\eta_0}\br)^{-\frac{1}{1-n}} \propto \sigma_0 \bl(\frac{k}{\alpha_0 M_C}\br)^{-\frac{1}{n}},
\eq
where $\sigma_0 \equiv \frac{H_0}{M_C}$ and $H_0=H(\eta_0)$ is the
characteristic Hubble expansion rate during inflation.
In our model $M_C=M_P$ and $\eta_k = \etap$, thus in the notation of \cite{Martin:2003kp} the correction to
the power spectrum
has an amplitude $\propto \sigma_0^2 \Bl(\frac{k}{\alpha_0
M_P}\Br)^{-\frac{2}{n}}\bl(C_1-\frac{C}{1+Q^{-2}}\br)$.
Although the effect is suppressed by $\sigma_0^2$, it is very
sensitive to the value of the deformation parameter. Thus, one can
significantly enhance the magnitude of the correction by taking
$Q$ to be arbitrarily close to $0$.

\section{A Special Case: de Sitter Inflation}
As a special case of power-law inflation, we now consider the de Sitter background.
In terms of conformal time, the
scale factor is given by $a(\eta)=-\frac{1}{H\eta}$, where $H \equiv \frac{\dot{a}}{a}$ is constant.
In region I the wavefunction for $|k\eta|$ large is
\beq
u_k^I = e_Q^{-ik\eta} \Bl(1 + \frac{1}{(k\eta)^2} \frac{[-1][-2]}{1-Q^{-4}} + \ldots
\Br),
\eq
where we are imposing initial conditions so that the mode has only
positive-frequency components.

In region II, the full mode
\beq
u^{II}_{k}=\sqrt{\eta} \Bl(A_k^{II} H_\nu(-k\eta)+B_k^{II} H^\ast_\nu(-k\eta)\Br),
\eq
with $\nu=\frac{3}{2}$ for de Sitter space, reduces to, in the large $|k\eta|$ limit,
\beq \label{uIIasympdS}
u^{II}_k= - \sqrt{\frac{2}{\pi |k\eta|}}
\Biggl\{ A_k e^{-ik\eta}\Bl(1 - \frac{i}{k\eta}\Br) + B_k e^{ik\eta}\Bl(1+\frac{i}{k\eta}\Br)
+{\cal O}\Bl(\frac{1}{(k\eta)^3}\Br) \Biggr\}.
\eq
Thus, we see that the de Sitter calculations match the power-law
inflation case provided $C_1=1$, $C_2=0$, $\phi=-\pi$ and
$b_2=\frac{[-1][-2]}{1-Q^{-4}}$.
The gravitational power spectrum for de Sitter inflation is, then,
\beq
P_g^{new}(k) = P_g^{st}(k) \: \Biggl\{1-\frac{1}{(k\etap)^2}\Bl(1-\frac{[-1][-2]}{1+Q^{-2}}\Br)
cos(2k\etap)\Biggr\}.
\eq
From $p=\frac{k}{a}=M_P$ and $a(\etap)=-\frac{1}{H\etap}$, one
finds $k\etap=-\frac{M_P}{H}$.
Rewriting the spectrum, we find
\beq
P_g^{new}(k) = P_g^{st}(k) \: \Biggl\{1-\Bl(\frac{H}{M_P}\Br)^2\Bl(1-\frac{[-1][-2]}{1+Q^{-2}}\Br)
cos\Bl(\frac{2M_P}{H}\Br)\Biggr\}.
\eq
The correction is scale invariant, as expected, and of order $\bl(\frac{H}{M_P}\br)^2$.
However, as in the case of power-law inflation, the deviation from
the standard spectrum is sensitive to the value of $Q$. By taking $Q$ to be small one can
substantially increase the size of the correction.
When $Q=1$ we recover the usual result, $P_g^{new}(k) \rightarrow P_g^{st}(k)$.

\section{Conclusions}

In this work we have analyzed the dependence of the spectrum of cosmological
perturbations on a quantum deformation of the wave equation assumed to hold
above some cutoff scale $\Lambda$, which we take to be $M_P$.
After finding the form of the $Q$-deformed mode, we have
imposed that, at the time of its creation, it contains only positive-frequency components.
This corresponds to choosing a Bunch-Davies-like vacuum.
We have expressed the standard mode solution, valid below $\Lambda$,
in terms of the $Q$-deformed solution by imposing appropriate matching conditions at the cutoff.
This has allowed us to calculate the spectrum of tensor and
scalar fluctuations for the case of power-law inflation.
The de Sitter background has been considered as a special
case of the latter. The power spectrum in this case is
defined only for tensor perturbations.

\begin{figure}[!]
\includegraphics[totalheight=6.5 in,keepaspectratio]{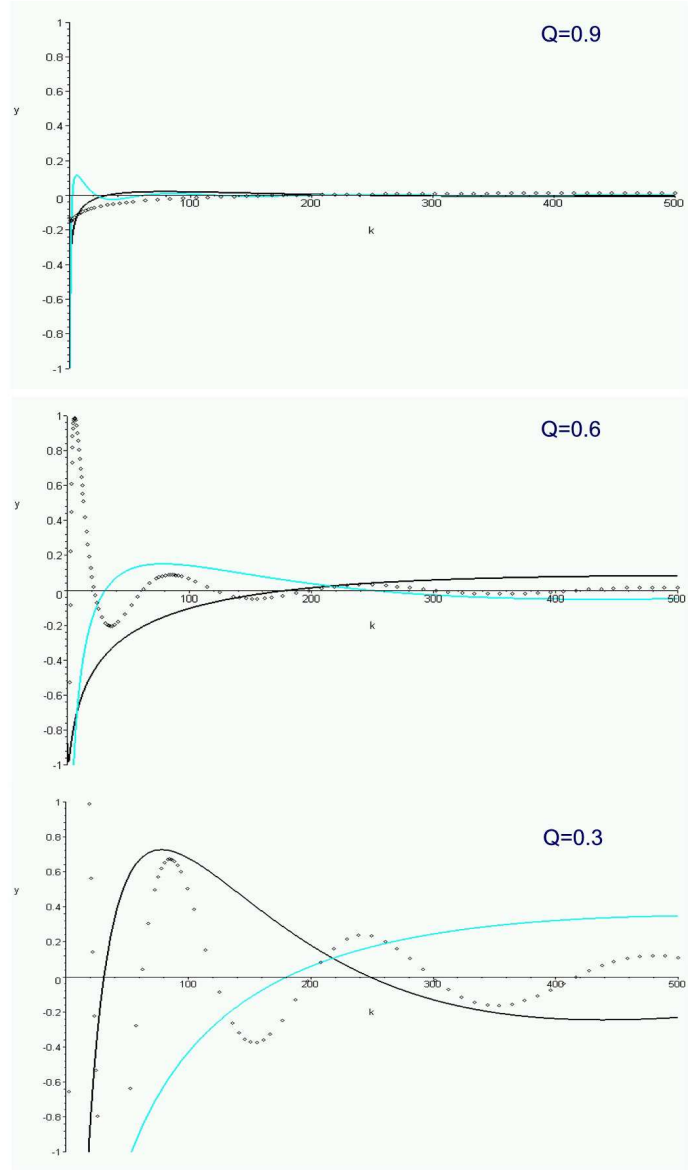}
\caption{Above we show the deviation from the standard
spectrum for different values of $Q$ and $n$.
In each plot the dark line corresponds to
$n=2$, the light one to $n=3$ and the dotted line to $n=4$.
We have set $\alpha_0=M_P=1$.
The effect of the $Q$-deformation is more pronounced in the small $k$ region.
Also, the size of the deviation increases significantly as $Q$ approaches $0$.}
\label{fig1}
\end{figure}

For power-law inflation the new spectrum has nonstandard dependence on comoving momentum,
while the de Sitter spectrum is scale invariant as expected.
In both inflationary scenarios the correction to the standard result is
characterized by an oscillatory pattern; this modulation has been argued to be a
general feature of the `minimal' models described in the introduction.
As shown in Fig.(\ref{fig1}), the amplitude of the oscillatory
pattern is strongly dependent on the deformation parameter.
Although only a few values of $Q$ were considered, it is apparent
from the plots that the size of the correction increases
as $Q$ gets farther away from the classical $Q=1$ limit.

Some of the minimal models \cite{Danielsson:2002kx,Easther:2002xe}
yield $\frac{H_0}{M_P}$ corrections, with $H_0$ the characteristic Hubble
parameter during inflation. When the deformation parameter is
very close to its classical value, $Q \sim 1$, the effect
of the quantum deformation is of order $\bl(\frac{H_0}{M_P}\br)^2$, significantly small.
However, values of $Q$ close to zero yield a considerable increase in the size of the
correction. If in fact $Q$ were small enough to overcome the
$\bl(\frac{H_0}{M_P}\br)^2$ suppression, the effect of the quantum
deformation could be measurable in the near future.
On the other hand, $Q$ near unity would yield a $\bl(\frac{H_0}{M_P}\br)^2$ correction to the
standard spectrum, most likely impossible to detect.

Since they predict a departure from the standard spectrum
of fluctuations, $Q$-deformations can in principle provide us with a window on
trans-Planckian physics. Whether the effect can be measured or not,
however, depends on the magnitude of the deformation.
Regardless, we have shown that $Q$-deformations offer an additional scenario
for physics above the Planck scale.


\begin{acknowledgments}
I would like to thank A. Jevicki and R. Brandenberger for assistance throughout
this work. I would also like to thank G. L. Alberghi, S. Ramgoolam and S. Watson for useful discussions.
\end{acknowledgments}


\begin{thebibliography}{99}

\bibitem{Martin:2000xs}
J.~Martin and R.~H.~Brandenberger,
Phys.\ Rev.\ D {\bf 63}, 123501 (2001) [arXiv:hep-th/0005209].

\bibitem{Brandenberger:2000wr}
R.~H.~Brandenberger and J.~Martin,
Mod.\ Phys.\ Lett.\ A {\bf 16}, 999 (2001)
[arXiv:astro-ph/0005432].

\bibitem{Amelino-Camelia:1999zc}
G.~Amelino-Camelia,
Lect.\ Notes Phys.\  {\bf 541}, 1 (2000) [arXiv:gr-qc/9910089].

\bibitem{Niemeyer} J.~C.~Niemeyer, Phys. Rev. {\bf D63},
123502 (2001), [arXiv:astro-ph/0005533];
J.~C.~Niemeyer, [arXiv:astro-ph/0201511].

\bibitem{Mersini:2001su}
L.~Mersini, M.~Bastero-Gil and P.~Kanti,
Phys.\ Rev.\ D {\bf 64}, 043508 (2001) [arXiv:hep-ph/0101210].

\bibitem{Starobinsky:2001kn}
A.~A.~Starobinsky,
Pisma Zh.\ Eksp.\ Teor.\ Fiz.\  {\bf 73}, 415 (2001)
[JETP Lett.\  {\bf 73}, 371 (2001)]
[arXiv:astro-ph/0104043].

\bibitem{Brandenberger:2002hs}
R.~H.~Brandenberger and J.~Martin,
Int.\ J.\ Mod.\ Phys.\ A {\bf 17}, 3663 (2002)
[arXiv:hep-th/0202142].

\bibitem{BM1} J.~Martin and R.~H.~Brandenberger, Proceedings of the Ninth
Marcel Grossmann Meeting on General Relativity, edited by
R.~T.~Jantzen, V.~Gurzadyan and R.~Ruffini, World Scientific,
Singapore, 2002, [arXiv:astro-ph/0012031].

\bibitem{NP2} J.~C.~Niemeyer and R.~Parentani, Phys. Rev. {\bf
D64}, 101301 (2001), [arXiv:astro-ph/0101451].

\bibitem{LLMU} M.~Lemoine, M.~Lubo, J.~Martin and J.~P.~Uzan,
Phys.~Rev.~{\bf D65}, 023510 (2002), [arXiv:hep-th/0109128].

\bibitem{BJM}
R.~H.~Brandenberger, S.~E.~Joras and J.~Martin,
Phys.\ Rev.\ D {\bf 66}, 083514 (2002), [arXiv:hep-th/0112122].

\bibitem{Kempf}
A.~Kempf, Phys. Rev. {\bf D63} (2001) 083514, [arXiv:astro-ph/0009209].

\bibitem{CGS}
C.~S.~Chu, B.~R.~Greene and G.~Shiu, Mod. Phys. Lett. {\bf A16}, 2231
(2001), [arXiv:hep-th/0011241].

\bibitem{EGKS}
R.~Easther, B.~R.~Greene, W.~H.~Kinney and G.~Shiu, Phys. Rev.
{\bf D64}, 103502 (2001), [arXiv:hep-th/0104102];
R.~Easther, B.~R.~Greene, W.~H.~Kinney and G.~Shiu, [arXiv:hep-th/0110226].

\bibitem{KempfN}
A.~Kempf and J.~C.~Niemeyer, Phys. Rev. {\bf D64}, 103501
(2001), [arXiv:astro-ph/0103225].

\bibitem{HS}
S.~F.~Hassan and M.~S.~Sloth,
[arXiv:hep-th/0204110].

\bibitem{Lizzi:2002ib}
F.~Lizzi, G.~Mangano, G.~Miele and M.~Peloso,
JHEP {\bf 0206}, 049 (2002), [arXiv:hep-th/0203099].

\bibitem{Brandenberger:2002nq}
R.~Brandenberger and P.~M.~Ho,
Phys.\ Rev.\ D {\bf 66}, 023517 (2002), [AAPPS Bull.\
{\bf 12N1}, 10 (2002)], [arXiv:hep-th/0203119].

\bibitem{Huang:2003zp}
Q.~G.~Huang and M.~Li,
[arXiv:hep-th/0304203].

\bibitem{Danielsson:2002kx}
U.~H.~Danielsson,
Phys.\ Rev.\ D {\bf 66}, 023511 (2002), [arXiv:hep-th/0203198].

\bibitem{Danielsson:2002qh}
U.~H.~Danielsson,
JHEP {\bf 0207}, 040 (2002), [arXiv:hep-th/0205227].

\bibitem{Danielsson:2002mb}
U.~H.~Danielsson,
JHEP {\bf 0212}, 025 (2002)
[arXiv:hep-th/0210058].

\bibitem{Goldstein:2002fc}
K.~Goldstein and D.~A.~Lowe,
Phys.\ Rev.\ D {\bf 67}, 063502 (2003), [arXiv:hep-th/0208167].

\bibitem{Alberghi:2003am}
G.~L.~Alberghi, R.~Casadio and A.~Tronconi,
[arXiv:gr-qc/0303035].

\bibitem{Armendariz-Picon:2003gd}
C.~Armendariz-Picon and E.~A.~Lim,
arXiv:hep-th/0303103.

\bibitem{Shenker}
N.~Kaloper, M.~Kleban, A.~Lawrence, S.~Shenker,
[arXiv:/hep-th/0201158].

\bibitem{Easther:2002xe}
R.~Easther, B.~R.~Greene, W.~H.~Kinney and G.~Shiu,
Phys.\ Rev.\ D {\bf 66}, 023518 (2002)
[arXiv:hep-th/0204129].

\bibitem{Niemeyer:2002kh}
J.~C.~Niemeyer, R.~Parentani and D.~Campo,
Phys.\ Rev.\ D {\bf 66}, 083510 (2002)
[arXiv:hep-th/0206149].

\bibitem{Martin:2003kp}
J.~Martin and R.~H.~Brandenberger,
arXiv:hep-th/0305161.



\bibitem{Majid:1990gq}
S.~Majid,
Int.\ J.\ Mod.\ Phys.\ A {\bf 5}, 4689 (1990).

\bibitem{Majid:1996nt}
S.~Majid,
arXiv:q-alg/9701001.

\bibitem{Teschner:1999my}
J.~Teschner,
Phys.\ Lett.\ B {\bf 458}, 257 (1999)
[arXiv:hep-th/9902189].

\bibitem{Mukhanov:1990me}
V.~F.~Mukhanov, H.~A.~Feldman and R.~H.~Brandenberger,
Phys.\ Rept.\  {\bf 215}, 203 (1992).

\bibitem{Jevicki:1998rr}
A.~Jevicki and S.~Ramgoolam,
JHEP {\bf 9904}, 032 (1999)
[arXiv:hep-th/9902059].

\bibitem{Mottola:ar}
E.~Mottola,
Phys.\ Rev.\ D {\bf 31}, 754 (1985).

\bibitem{Allen:ux}
B.~Allen,
Phys.\ Rev.\ D {\bf 32}, 3136 (1985).



\end{thebibliography}
\end{document}